\documentclass[%
 aip,
 amsmath,amssymb,
 reprint,%
]{revtex4-1}

\usepackage{graphicx}
\usepackage{dcolumn}
\usepackage{bm}

\usepackage[utf8]{inputenc}
\usepackage[T1]{fontenc}
\usepackage{mathptmx}
\usepackage{color}

\begin{document}

\newcommand{\etal}{\emph{et al.\/}\xspace}
\newcommand{\ie}{i.e.\ }
\newcommand{\eq}[1]{Eq.\,(\ref{#1})\xspace}

\preprint{AIP/123-QED}

\title[Improving cold-atom sensors with quantum entanglement: Prospects and challenges]{Improving cold-atom sensors with quantum entanglement: Prospects and challenges}


\author{Stuart~S.~Szigeti}
\email{stuart.szigeti@anu.edu.au}
\affiliation{ 
Department of Quantum Science, Research School of Physics, The Australian National University, Canberra 2601, Australia.
}%
\author{Onur Hosten}%
\email{onur.hosten@ist.ac.at}
\affiliation{Institute of Science and Technology Austria, 3400 Klosterneuburg, Austria.
}%
\author{Simon~A.~Haine}
\email{Simon.Haine@anu.edu.au}
\affiliation{ 
Department of Quantum Science, Research School of Physics, The Australian National University, Canberra 2601, Australia.
}%

\date{\today}

\begin{abstract}
Quantum entanglement has been generated and verified in cold-atom experiments and used to make atom-interferometric measurements below the shot-noise limit. However, current state-of-the-art cold-atom devices exploit separable (i.e. unentangled) atomic states. This Perspective piece asks the question: can entanglement usefully improve cold-atom sensors, in the sense that it gives new sensing capabilities unachievable with current state-of-the-art devices? We briefly review the state-of-the-art in precision cold-atom sensing, focussing on clocks and inertial sensors, identifying the potential benefits entanglement could bring to these devices, and the challenges that need to be overcome to realize these benefits. We survey demonstrated methods of generating metrologically-useful entanglement in cold-atom systems, note their relative strengths and weaknesses, and assess their prospects for near-to-medium term quantum-enhanced cold-atom sensing. 
\end{abstract}

\maketitle
Atom interferometry is a leading precision measurement technology that harnesses the wave-like interference of atoms to make precise measurements of time~\cite{Nicholson:2015}, accelerations~\cite{Canuel:2006}, rotations~\cite{Gustavson:1997}, gravity~\cite{Peters:2001}, gravity gradients~\cite{Snadden:1998}, magnetic fields~\cite{Vengalattore:2007}, the fine structure constant~\cite{Parker:2018}, and Newton's gravitational constant~\cite{Rosi:2014}. Future applications of atom interferometry include inertial navigation~\cite{Jekeli:2005, Battelier:2016, Cheiney:2018}, mineral exploration and recovery~\cite{vanLeeuwen:2000,Evstifeev:2017}, groundwater monitoring~\cite{Canuel:2018}, satellite gravimetry~\cite{Tino:2013, Carraz:2014,Chiow:2016, Douch:2018, Abrykosov:2019, Migliaccio:2019, Trimeche:2019}, and space-based experiments that test general relativity and candidate theories of quantum gravity~\cite{Dimopoulos:2007, Aguilera:2014,Williams:2016,Becker:2018, Tino:2020}. These applications require a new generation of atom interferometers capable of highly precise, stable measurements in compact, low-weight configurations, that can also operate in real-world field conditions~\cite{Farah:2014,Freier:2016,Bidel:2018,Grotti:2018,Bongs:2019}. Consequently, improvements to cold-atom sensors are not solely aimed at improving precision~\cite{Geiger:2020}: increased stability~\cite{Hu:2013,Gillot:2014,Freier:2016,Menoret:2018}, increased accuracy~\cite{Karcher:2018}, increased dynamic range~\cite{Lautier:2014}, increased measurement rate~\cite{Rakholia:2014}, and decreased size, weight, and power (SWaP)~\cite{Zoest:2010,Hinton:2017,Wigley:2019} are all desirable traits, alongside improved performance in the presence of technical and environmental noise (e.g. due to vehicle motion).

Quantum entanglement offers a promising route to improved atom interferometry, since certain entangled atomic states allow relative phase measurements below the shot-noise limit (SNL) (the ultimate sensitivity limit achievable by uncorrelated sources)~\cite{Giovannetti:2006}. Such \emph{quantum-enhanced atom interferometry} could be beneficial when the atom number of the atomic source cannot be increased further due to technical issues or operational requirements. Spin-squeezed states~\cite{Kitagawa:1993} have been the focus of most experimental quantum-enhanced atom interferometry research to date, since they are relatively easy to generate, characterize, and directly incorporate into existing atom interferometry schemes. The first experimental demonstrations of metrological spin squeezing in cold atoms occurred little more than a decade ago~\cite{Esteve:2008,Appel:2009,Riedel:2010,Gross:2010,Schleier-Smith:2010b}, and were quickly followed by proof-of-principle sub-shot-noise atom interferometric measurements~\cite{Leroux:2010b,Lucke:2011}. Since then, tremendous scientific and technical progress has been made both in the size of metrological gain and the total number of entangled particles~\cite{Pezze:2018}, cementing cold atoms as the superior platform for generating and manipulating metrologically-useful nonclassical states. To date, an ensemble of $10^6$ cold atoms has provided the largest degree of spin squeezing (reduced spin variance of 20~dB), directly resulting in the most precise relative phase measurement ever made with cold atomic matter-waves~\cite{Hosten:2016}.

Despite these successes, quantum-correlated atomic sources have not yet provided a useful improvement to high precision cold-atom sensing~\footnote{Although quantum-enhanced atomic magnetometers have demonstrated sensitivities at particular spatial resolutions close to state-of-the-art, they have not yet provided a new sensing capability unachievable by magnetometers that do not exploit quantum entanglement.}. A quantum-enhanced (sub-shot-noise) atom interferometer has never made a measurement of a physical quantity that outperforms a state-of-the-art measurement with uncorrelated particles. Indeed, no quantum-enhanced atom interferometer has demonstrated \emph{any} sensitivity to inertial quantities, even in laboratory-based proof-of-principle apparatus. Neither has quantum entanglement improved sensing capabilities via increased stability, increased bandwidth, or decreased SWaP. This is understandable; as we describe below, it is extremely challenging to incorporate quantum entanglement into a high precision atom interferometer such that both the entanglement and high precision of the measurement device are preserved. This is true across all quantum technology platforms, not just cold atoms. Arguably, Advanced LIGO is the only useful quantum-enhanced measurement device in existence, since the injection of 3~dB of squeezed light significantly increased the detection rate of gravitational wave candidates~\cite{Acernese:2019,Tse:2019}. Nevertheless, the potential benefits to cold-atom sensing offered by quantum entanglement make overcoming these challenges an extremely worthwhile goal.

This Perspective outlines the near-to-medium term prospects for useful quantum-enhanced measurements with cold atoms, focussing on clocks and inertial sensing. We review the current state-of-the-art in cold-atom sensing, explicitly identifying what is needed to \emph{be} state-of-the-art and how entanglement could be used to improve current cold-atom technology. We consider methods of generating metrologically-useful entanglement for atomic clocks, inertial sensors, and magnetometers, and
assess their suitability for usefully improving cold-atom sensing. 

\section{Sensitivity limits for cold-atom sensors}
\begin{figure}[t]
	\begin{center}
		\includegraphics[width=\columnwidth]{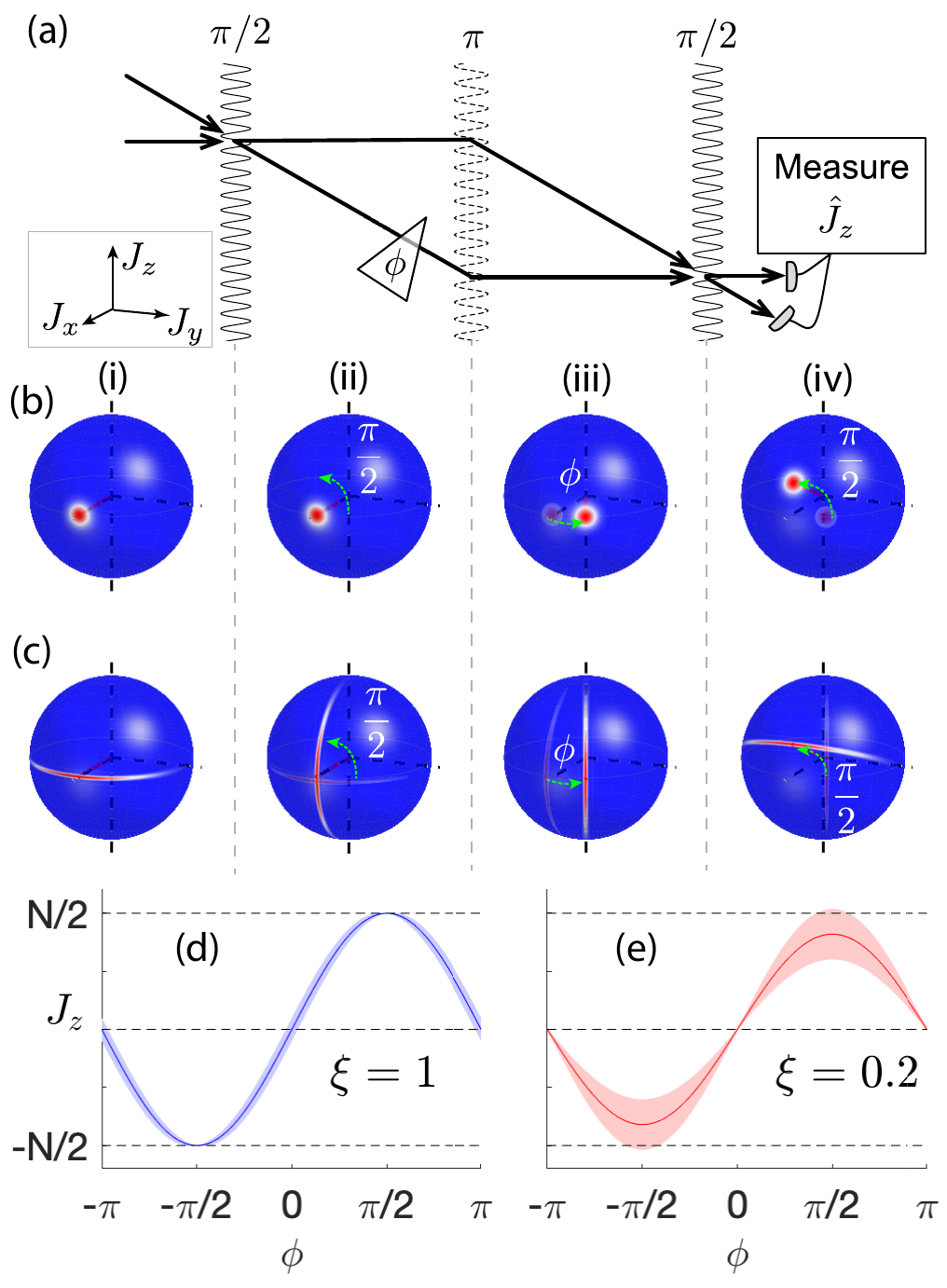}
		\caption{(a) Schematic illustrating how a relative phase $\phi$ between two atomic matter-wave modes can be measured. An initial two-mode system of $N$ atoms, which could be entangled, is coherently beamsplit ($\pi/2$ pulse), allowed to evolve for a period of interrogation, and then subsequently recombined by a second beamsplitting pulse prior to measuring the atom-number difference between the two modes. If the two modes have different momenta, a $\pi$ pulse (mirror) can redirect the states such that they overlap at the final beamsplitter. (b) Bloch spheres showing the evolution of a CSS's Wigner quasiprobability distribution \cite{Dowling:1994} through the interferometer depicted in (a). Since a CSS input is unentangled, $\xi = 1$. The axes of the Bloch spheres are defined in the inset. (c) As in (b), except for an initial spin-squeezed state with minimum variance in $J_z$ and a spin-squeezing parameter of $\xi = 0.2$. (d) For uncorrelated particles, such as the CSS input state shown in (b,i), a measurement of $J_z$ gives an average signal (solid line) that varies sinusoidally with phase $\phi$. However, shot noise on this signal (shaded region) limits the phase measurement to no better than $\Delta \phi = 1/\sqrt{N}$. (e) A correlated source of particles, such as the input spin-squeezed state shown in (c,i), allows a phase measurement at sensitivity better than $1/\sqrt{N}$. Here, $\Delta \phi = \xi / \sqrt{N}$ with $\xi = 0.2$, a factor of five below the SNL.}
		\label{fig:signal_and_noise}
	\end{center}
\end{figure}

Like their optical counterparts, atom interferometers rely upon the principle of wave interference; the coherent sum of two atomic matter-waves gives an outcome that depends upon their relative phase. Typically, an electromagnetic pulse splits an initial atomic ensemble into a coherent superposition of two states (internal and/or motional). For clocks these two states are magnetically insensitive internal states, and for inertial sensors these are distinct spatial paths. Each component of the superposition then acquires a phase during a subsequent period of free evolution (called the interrogation time). A final beamsplitting pulse then coherently interferes the two states, resulting in an output that contains information about the \emph{difference} in the accumulated phase, $\phi$. For the interferometer to be useful, this relative phase difference needs to depend upon a physical quantity of interest. By measuring some observable $\hat{S}$ at the interferometer output, $\phi$ can be determined to a precision
\begin{equation}
    \Delta \phi = \frac{\sqrt{\text{Var}(\hat{S})}}{\left| \partial \langle \hat{S} \rangle / \partial \phi \right|} \equiv \frac{\xi}{\sqrt{N}} \label{Delta_phi},
\end{equation}
where the parameter
\begin{equation}
    \xi = \sqrt{N} \frac{\sqrt{\text{Var}(\hat{S})}}{\left| \partial \langle \hat{S} \rangle / \partial \phi \right|} \label{xi_defn}
\end{equation}
quantifies the effect of quantum noise and fringe contrast on the sensitivity, with $N$ the number of atoms involved in the measurement. A commonly measured observable is the population difference between the two states, yielding a signal that varies sinusoidally with the relative phase (see Fig.~\ref{fig:signal_and_noise}).

Irrespective of the nature of the two states, mapping the problem onto the physics of a collection of spin-1/2 particles clarifies the analysis both mathematically and pictorially~\cite{Yurke:1986}. For a fixed number of identically-addressed atoms, the system that defines our atom interferometer is formally equivalent to a collective spin $\bm{\hat{J}} = (\hat{J}_x, \hat{J}_y, \hat{J}_z)$ of length $|\langle \bm{\hat{J}} \rangle | = N/2$. Here $\hat{J}_z$ is half the population difference of the two states and $\hat{J}_x$ and $\hat{J}_y$ encode coherences between the states. A population-difference measurement is therefore equivalent to a $\hat{J}_z$ measurement. For an ensemble of $N$ uncorrelated atoms, each initialized in the same state, the collective state is a coherent spin state (CSS). Taking a CSS with spin projection $J_z=-N/2$ as the interferometer input gives an output state with $\langle \hat{J}_z \rangle = \tfrac{N}{2} \cos \phi$ and $\text{Var}(\hat{J}_z) = \frac{N}{4}\sin^2 \phi $. Therefore, by biasing the interferometer to operate around $\phi = \pi/2$, where it is most sensitive to a change in $\phi$, a measurement $\hat{S} = \hat{J}_z$ yields $\xi = 1$, and therefore can resolve a relative phase shift no better than the atomic SNL $\Delta \phi = 1/\sqrt{N}$ [see Fig.~\ref{fig:signal_and_noise}].

However, certain quantum states that are not simple product states of each particle (i.e. the particles are entangled) allow for $\xi <1$ and therefore relative phase measurements that surpass the SNL. The most common quantum enhancement is spin squeezing, which reduces the variance of the collective spin below $N/4$ along some axis. For a spin-squeezed state with minimum variance in $\hat{J}_z$ and a maximum average spin length in the $\hat{J}_x$ direction, $\xi$ reduces to the Wineland spin-squeezing parameter~\cite{Wineland:1994}
\begin{equation}
     \xi_\text{ssp} = \sqrt{N} \frac{\sqrt{\text{Var}(\hat{J}_z)}}{|\langle \hat{J}_x\rangle|}. \label{Delta_phi_spinsqz}
\end{equation}

Observing $\xi < 1$ implies that there is entanglement between the atoms~\cite{Sorensen:2001,Giovannetti:2006}. However, there is a broader class of entangled states that allow sub-SNL interferometry, termed \emph{metrologically-useful entangled states}, characterized by $F_Q > N$, where $F_Q$ is the quantum Fisher information (QFI)~\cite{Pezze:2009,Toth:2012,Hyllus:2012}. The QFI places a fundamental lower bound on the achievable phase sensitivity, called the quantum Cram{\'e}r-Rao bound~\cite{Braunstein:1994}: $\Delta \phi \geq 1 / \sqrt{F_Q}$. This makes the connection between sub-SNL metrology and QFI explicit, since metrologically-useful states (possessing $F_Q > N$) can achieve $\Delta \phi \leq 1/\sqrt{N}$. It also shows that entanglement between particles is needed to achieve sub-SNL sensitivities (since separable states have $F_Q \leq N$)~\cite{Giovannetti:2006}, and furthermore that not all entangled states are metrologically useful~\cite{Hyllus:2010}. Although the QFI is in general difficult to compute~\cite{Toth:2014,Pezze:2018}, for the interferometer depicted in Fig.~\ref{fig:signal_and_noise}(a) with a pure state input~\cite{Pezze:2009}, $F_Q = 4 \text{Var}(\hat{J}_y)$. This immediately shows that there are metrologically-useful entangled states that are not spin squeezed. For example, the Greenberger-Horne-Zeilinger (GHZ) state, which is a coherent superposition of all particles in one state and all particles in the other state, has $F_Q = N^2$ but does not display spin squeezing~\cite{Giovannetti:2004}. Although these so-called non-Gaussian entangled states in principle have superior metrological potential to easier-to-characterize entangled states such as spin-squeezed states, they generally require more complicated preparation and readout procedures, and are more susceptible to losses~\cite{Demkowicz-Dobrzanski:2012}. The first useful quantum-enhanced cold-atom sensors are therefore most likely to exploit entangled states that are easier to generate and can be characterized by low-order moments of the collective spin operators. This will therefore be the focus of this Perspective piece.

\section{Atomic clocks}
\subsection{Current state-of-the-art}
Atomic clocks exploit the frequency stability of long-lived energy levels in isolated atoms. They operate by comparing the frequency of a local oscillator (LO) to the frequency of a reference atomic transition, which is used to stabilize the ticking period of the LO. This is accomplished by Ramsey spectroscopy, where an electromagnetic pulse (microwave or optical) creates a superposition of two internal atomic states and, after a waiting period (interrogation time), interferes them and extracts a signal proportional to the phase drift $\phi$ between the LO and the reference transition (see Fig.~\ref{fig:signal_and_noise}(b) and (c)). 

Microwave atomic clocks that use the caesium hyperfine transition frequency of $\sim 9 \times 10^{9}$~Hz still remain the standard for the definition of time, permitting a measurement of time with a fractional uncertainty of one part in $10^{16}$. However, laboratory optical lattice clocks of strontium or ytterbium atoms, which exploit long-lived optical transitions with frequencies in the $10^{14}$~Hz range~\cite{Ludlow:2015}, have already surpassed their microwave counterparts, achieving a $10^{-18}$ level fractional frequency stability~\cite{Oelker:2019,Schioppo:2017,Bloom:2014}. 

The quantum-noise-limited fractional stability of an atomic clock is given by
\begin{equation}
    \sigma_\mathrm{clock}=\frac{1}{\omega_0 T} \sqrt{\frac{T_C}{\tau}}\frac{\xi}{\sqrt{N}}  \label{sigma_clock},
\end{equation}
where $\omega_0$ is the angular frequency of the clock transition, $T$ is the interrogation time (or Ramsey time; i.e. the duration that the atoms remain in a superposition), $T_C$ is the clock cycle time, $\tau$ is the total averaging time, and the last factor is the phase sensitivity for $N$ atoms given in Eq.~(\ref{Delta_phi}). This simplified expression for $\sigma_\mathrm{clock}$ ignores the Dick noise due to aliasing of the high-frequency LO noise when operating with non-unity duty cycle ($T<T_C$). Note that state-of-the-art atomic clocks can run with unity duty cycle~\cite{Schioppo:2017} ($T=T_C$), eliminating the dead time due to atomic state preparation and readout, further simplifying the stability expression to $\sigma_\text{clock}=(\omega_0\sqrt{T \tau})^{-1}(\xi/\sqrt{N})$. 

Microwave fountain clocks typically use around $10^6$ atoms~\cite{Guena:2014}, while optical lattice clocks use $\sim 10^3$ atoms, both limited by the systematic effects on the clock frequency due to interactions between the atoms. Interrogation times for microwave clocks reach a fraction of a second, limited by the free-fall time. It is possible to lift this limitation by magnetically trapping the atoms, which has been demonstrated at the $10^{-13}$ fractional frequency stability level~\cite{Szmuk:2015}. In this configuration, atom-atom interactions can actually play a positive role by increasing the coherence times up to the minute scale~\cite{Deutsch:2010}. Optical lattice clocks have already demonstrated interrogation times on the order of a second, and in principle could achieve interrogation times in the $\sim100$ second regime, limited by the excited state lifetime. Currently, the record systematic uncertainty demonstrated by an atomic clock is $9.4 \times 10^{-19}$, and is held by a single aluminium ion-based optical clock~\cite{Brewer:2019}. Nevertheless, with further unravelling of systematic effects, optical lattice clocks utilizing many atoms have the potential to significantly surpass single ion clocks.

Optical clocks offer more than direct technological improvements to, for example, the long-term stability of global navigation satellite systems. State-of-the-art optical clocks operate in a new territory where the ticking rate is appreciably affected by height-dependent gravitational potentials even at laboratory length scales ($\sim 10^{-16}$ per metre fractional frequency shift on Earth). This opens new doors to applications in geodesy~\cite{Bondarescu:2012} as well as precision tests of general relativity, provided that transportable atomic clocks can achieve the same precision obtained in laboratory experiments. Recent progress in a 450 m tower experiment with transportable clocks, for example, has led to tests of gravitational redshift at the $10^{-5}$ level~\cite{Takamoto:2020} with the clocks indeed operating at the $10^{-18}$ stability and accuracy level. For a more complete list of applications see Ref.~[\cite{Ludlow:2015}].

\subsection{Quantum-enhanced atomic clocks and potential benefits}
Based on Eq.~(\ref{sigma_clock}), given a fixed atomic transition frequency, the path to improved clock performance goes through (1) larger atom numbers - which is limited due to interactions, (2) longer interrogation times, and (3) operation beyond the SNL with quantum entanglement ($\xi<1$). Ideally, one would at least like to take advantage of the last two points; however, a natural question is whether they are compatible with each other. 

Squeezed microwave Rb atomic clocks operating inside an optical cavity have been demonstrated~\cite{Leroux:2010b,Hosten:2016}, showing that an entanglement-enhanced clock can average down to the same stability as a shot-noise-limited clock $\xi^{-2}$ times faster; a factor of ten in the best demonstrated case. Alternatively, the same results show that a given precision can be achieved with a factor of $\xi^2$ fewer atoms, opening up the possibility of building sensors with decreased SWaP for mobile applications. In the first demonstrations, the interrogation times were limited to several hundred microseconds, leaving open the question of wheather spin squeezing could be preserved at the longer time scales relevant to advanced atomic clocks. A full clock demonstration with squeezed states at the 10~ms time scale has been recently reported by releasing cavity-squeezed atoms into free fall and detecting them with fluorescence imaging~\cite{Malia:2020}. Most recently, an atom-chip-trap experiment has reported that metrologically-useful squeezing can be preserved for times up to one second~\cite{Huang:2020}, paving the way towards entanglement enhancement at the largest time scales used by microwave clocks. 

The first squeezed optical lattice clock was only demonstrated very recently~\cite{Pedrozo:2020}, achieving a metrological enhancement of 4.4~dB ($\xi = 0.60$); again using optical-cavity-based techniques. Specifically, the spin squeezing was generated in the ground state manifold of ytterbium ($^{171}$Yb) via cavity mediated atom-light interactions, which was then mapped onto the optically excited state, in effect shuffling the squeezing between observables. Although the interrogation times were limited to $\sim$1 ms, this development provides a viable route to quantum-enhanced state-of-the-art optical clocks.

The long-term (hour- to day-scale) stability of optical clocks are far from saturated. However, improving the absolute stability at these time scales requires extensive statistical characterization of systematic effects. Such characterizations can become extremely time consuming and less-and-less feasible. Quantum-enhanced clocks have the potential to significantly improve this process, cutting the required characterization times by achieving a given precision faster as described above.

\subsection{Requirements for quantum-enhanced atomic clocks}
To achieve the best performance, state-of-the-art clocks maximize the available interrogation time. In this mode of operation, the phase drift between the LO and the atomic states can grow to a value of order 1 rad (due to finite LO coherence time). Thus, at first sight, the use of entanglement may seem incompatible with long-interrogation-time clocks. The problem can be understood pictorially from Fig. 1(c): The more squeezed an initial state in the $J_z$ direction, the larger the anti-squeezing around the equator ($J_x$-$J_y$ plane). Due to the curvature of the Bloch sphere, as the unknown phase $\phi$ grows larger, the anti-squeezed noise in $J_x$ and $J_y$ couples into the $J_z$ projection, eventually rendering the state noisier than an unsqueezed one for population-difference measurements. This effect generated scepticism~\cite{Andre:2004,Leroux:2017,Braverman:2018}, questioning whether quantum entanglement can actually improve the absolute stability of atomic clocks. A recent careful analysis~\cite{Schulte2:2020}, which takes into account all relevant sources of instability (Dick effect, finite LO coherence and quantum projection noise), finds that in a standard single-ensemble Ramsey clock, there could in fact be room for an absolute performance enhancement due to spin squeezing for existing systems where atom numbers are on the smaller side.

Future atomic clocks could incorporate several improvements not considered by Ref.~[~\cite{Schulte2:2020}], making the prospect of a useful quantum-enhanced atomic clock even more promising than the analysis of Ref.~[~\cite{Schulte2:2020}] suggests. These improvements could push the total noise due to the Dick effect and the coupling of anti-squeezing below the SNL; this is a necessary operating condition for quantum-enhanced atomic clocks, since entanglement can only benefit atomic clocks limited by atomic shot noise (i.e. projection noise). In particular, it has been experimentally shown that the Dick effect can be completely eliminated by using two clock ensembles~\cite{Schioppo:2017}. Furthermore, there are two methods for overcoming the coupling of anti-squeezing, thereby increasing the interrogation times that still allow an enhancement due to spin squeezing. The first is the use of adaptive measurements on a single clock ensemble, and the second is the use of multiple clock ensembles with measurement and feed forward.

In the single-ensemble adaptive measurement method~\cite{Borregaard:2013}, a weak measurement of the collective spin state is followed by a feedback that rotates the state by an amount determined by the measurement outcome, aligning it with $J_z=0$ to within the measurement resolution. The weak nature of the measurement renders the amount of measurement backaction negligible, and the final state is left free of undesired noise mixing. At this point a strong measurement follows to readout the atomic states. Elements of this proposal were demonstrated with microwave transitions where random population shifts were imparted onto the atoms, weakly measured in a non-destructive fashion using an optical cavity, and corrected via follow-up microwave rotations~\cite{Vanderbruggen:2013}. Similar methods were also employed in the context of deterministic preparation of QND-based spin-squeezed states~\cite{Cox:2016}.

The possibility of extending interrogation times using multiple clock ensembles was first considered in the context of unentangled atoms~\cite{Borregaard2:2013,Rosenband:2013,Hume:2016}, with the goal of extending the interrogation times well beyond relative LO-clock phase drifts of 1 rad. In this scheme, multiple clock ensembles are interrogated simultaneously and the same LO is used for all ensembles. The result of the measurement of one clock is fed forward to the next longer-duration clock via a phase correction. This allows exponential improvements to interrogation times with the number of clock ensembles. A similar idea can extend the interrogation times of spin-squeezed atomic clocks. In this case, however, one tries to overcome not the 1 rad limit, but a much smaller relative LO-clock phase drift that appreciably contaminates the measurements with the anti-squeezed noise. Such a protocol was analyzed very recently~\cite{Pezze:2020}, where one squeezed clock was considered together with another unentangled clock. This analysis showed that feeding forward the measurement result of the unentangled ensemble to the squeezed ensemble allows entanglement enhancement to be maintained for longer durations in presence of LO noise. The phase of the unentangled ensemble could also be continuously tracked in such a proposal by locking the LO phase to the atomic phase via weak measurements~\cite{Shiga:2012}. The basic elements of this continuous tracking have already been demonstrated~\cite{Kohlhaas:2015}.

Research into quantum-enhanced atomic clocks provide important lessons and technological development that can also inform the design of more complicated atom interferometry schemes, including those relevant to cold-atom inertial sensors.

\section{Cold-atom inertial sensors}
\subsection{Current state-of-the-art}
\begin{figure}[t!]
	\begin{center}
		\includegraphics[width=0.7\columnwidth]{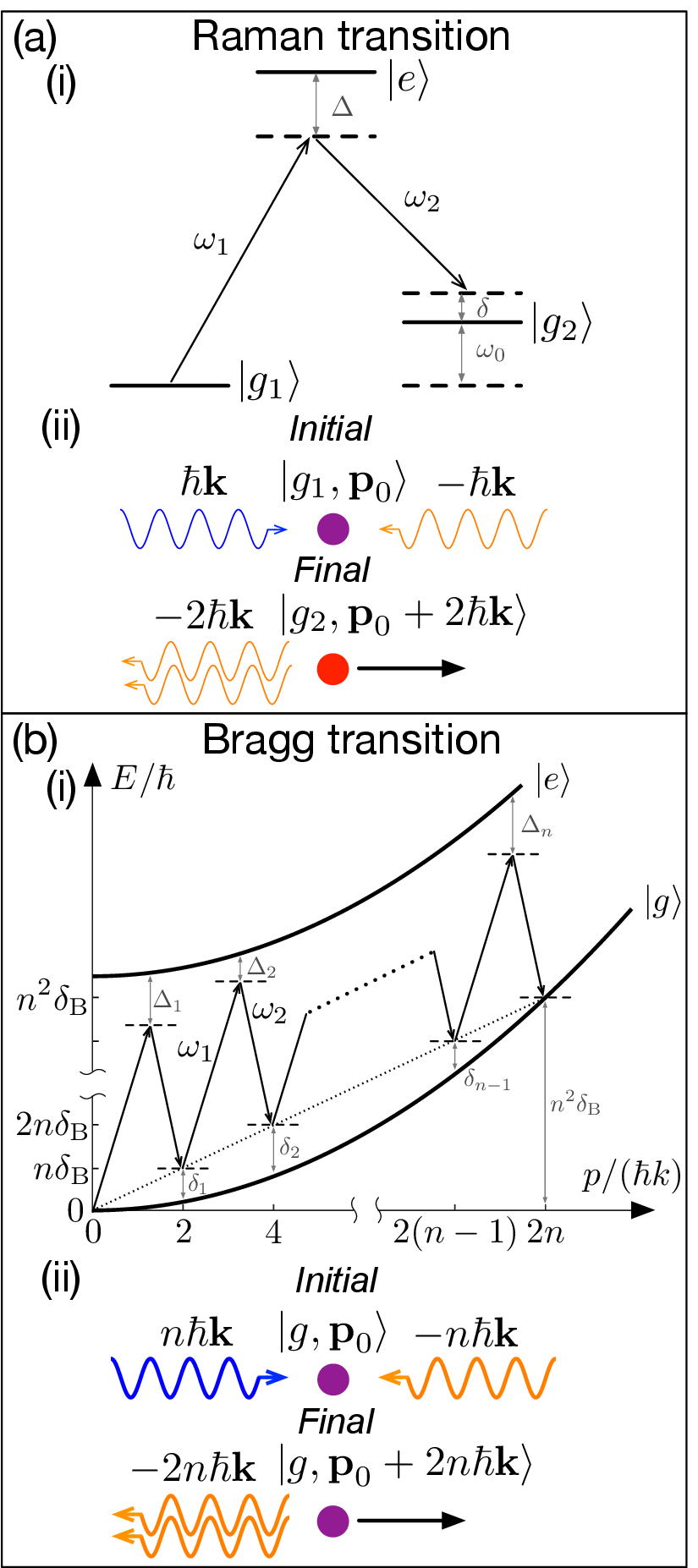}
		\caption{(a) Level diagram (i) and schematic (ii) for a two-photon Raman transition between $|g_1,\textbf{p}_0\rangle$ and $|g_2,\textbf{p}_0 + 2\hbar \textbf{k}\rangle$. Atoms with initial momentum $\textbf{p}_0$ are illuminated with light from two counter-propagating lasers of frequencies $\omega_1$ and $\omega_2$ and wavevectors $\textbf{k}$ and $-\textbf{k}$, respectively. $\Delta$ is the one-photon detuning from some excited state $|e \rangle$ and $\delta$ is the two-photon detuning, which depends upon the momentum of the atoms. When $\Delta \gg \delta$ and $\omega_1 - \omega_2 \approx \omega_0$, atomic population in $|g_1\rangle$ can be resonantly transferred to $|g_2\rangle$, accompanied by a $2 \hbar \textbf{k}$ momentum kick to the atoms, as shown in (a,ii). (b) Level diagram (i) and schematic (ii) for a $2n$-photon Bragg transition between $|g,\textbf{p}_0\rangle$ and $|g,\textbf{p}_0 + 2 n \hbar \textbf{k}\rangle$. The solid parabolic curves are the free atom dispersion relations between momentum $p$ and energy $E$ for the ground and excited internal states. Here $\delta_\text{B} = 4 \omega_r$, where $\omega_r = \hbar k^2 / (2m)$ is the recoil frequency ($k = |\textbf{k}|$), and the two-photon detunings are $\delta_m = 4 m (n-m) \omega_r$. Provided the one-photon detunings $\Delta_m$ are much larger than $\delta_m$ and $\omega_1 - \omega_2 \approx n \delta_\text{B}$, a $2n$-photon transition couples momentum eigenstates $|\textbf{p}\rangle$ and $|\textbf{p} + 2 n \hbar \textbf{k}\rangle$ via $2n-1$ intermediate states (horizontal dashed lines). These intermediate states are negligibly populated provided the effective $2n$-photon Rabi frequency is small compared to $\delta_m$~\cite{Muller:2008}.}
		\label{fig:Raman_Bragg_figure}
	\end{center}
\end{figure}

Cold-atom inertial sensing requires (1) a coherent matter-wave source and (2) the ability to coherently beam split, reflect, and recombine \emph{motional} atomic states. This latter property separates atomic clocks from cold-atom inertial sensors, since an acceleration or rotation measurement requires a space-time separation of the matter-waves that form the two interferometer arms~\cite{Kritsotakis:2018}. Current state-of-the-art cold-atom accelerometers and gyroscopes effect beam splitting with standing waves of light, formed via two counter-propagating laser pulses tuned to drive Raman~\cite{Kasevich:1991} or Bragg~\cite{Altin:2013} transitions in the atoms (see Fig.~\ref{fig:Raman_Bragg_figure}). In the standard Mach-Zehnder configuration where the atoms are in free fall, a uniform acceleration $\textbf{a}$ induces a relative phase shift between the two interferometer arms~\cite{Kasevich:1991}: $\phi = \textbf{k} \cdot \textbf{a} T^2$, where $\hbar \textbf{k}$ is the momentum imparted by the beamsplitters and mirrors and $T$ is the time between pulses (interrogation time). This phase can be extracted via a population-difference ($\hat{J}_z$) measurement at the interferometer output, yielding an acceleration measurement with per-shot sensitivity
\begin{equation}
	\Delta a = \frac{\xi}{\sqrt{N} k_\parallel T^2}, \label{Delta_a}
\end{equation}
where $N$ is the atom number,  $k_\parallel$ is the component of $\textbf{k}$ parallel to the acceleration $\textbf{a}$, and $\xi$ is defined in Eq.~(\ref{xi_defn}). Similarly, a rotation at rate $\bm{\Omega}$ induces the Sagnac phase shift~\cite{Gustavson:2000} $\phi = 2 m \bm{\Omega} \cdot \textbf{A} / \hbar$, where $m$ is the atomic mass and the magnitude of $\textbf{A}$ is the physical area enclosed by the atom interferometer. A $\hat{J}_z$ measurement at the interferometer output yields a measurement of rotation rate with a per-shot sensitivity
\begin{equation}
	\Delta \Omega = \frac{\xi \hbar }{\sqrt{N} 2 m A_\parallel}, \label{Delta_Omega}
\end{equation}
where $A_\parallel$ is the component of $\textbf{A}$ parallel to $\bm{\Omega}$ \cite{Anandan:1981, Cronin:2009, Haine:2016b}.

Cold-atom inertial sensors have demonstrated absolute gravity measurements at sensitivities of one part in $10^{8} -- 10^{9}$ in both fixed laboratory environments~\cite{Peters:1999,Peters:2001,Muller:2008c,Hardman:2016} and on mobile platforms~\cite{Farah:2014,Freier:2016,Menoret:2018}. These outperform the sensitivities of the most precise falling corner-cube devices by between a factor of 2--3.~\cite{Niebauer:1995,Peters:2001,Gillot:2014,Freier:2016} Differential measurements between two absolute cold-atom gravimeters have reported sensitivities approaching $10^{-9} g / \sqrt{\text{Hz}}$~\cite{McGuirk:2002,Rosi:2014}, allowing precision measurements of the Earth's gravity gradient~\cite{Snadden:1998,DAmico:2016,Wu:2019} and tests of the weak equivalence principle with quantum test particles~\cite{Schlippert:2014,Zhou:2015,Rosi:2017}. Cold-atom acceleration sensors have demonstrated impressive short-term sensitivities and stability onboard aircraft~\cite{Geiger:2011} and a ship~\cite{Bidel:2018}, with promising prospects for future deployment in space~\cite{Becker:2018}.

Atomic matter-wave gyroscopes have achieved short-term sensitivities below $10^{-9}$ rad/s/$\sqrt{\textrm{Hz}}$ in laboratory-based experiments~\cite{Gustavson:2000,Durfee:2006}, which is comparable to the short-term sensitivities of typical navigation-grade ring-laser and fibre-optical gyroscopes~\cite{Garrido-Alzar:2019}. In contrast, compact cold-atom gyroscopes have demonstrated short-term sensitivities closer to $10^{-5}$~rad/s/$\sqrt{\textrm{Hz}}$.~\cite{Canuel:2006} Ring-laser and fibre-optical gyroscopes have additional advantages over current cold-atom devices, including high bandwidth, high dynamic range, low power and cost effectiveness. However, the true benefit of cold-atom sensors lies in their extremely low in-run and scale-factor bias drifts, which in principle could be orders of magnitude lower than mechanical and optical gyroscopes~\cite{Garrido-Alzar:2019}. To date, the most stable cold-atom gyroscope achieved a stability of $3 \times 10^{-10}$ rad/s, comparable to strategic-grade fibre-optic gyroscopes~\cite{Savoie:2018}. The long-term stability of cold-atom inertial sensors also enables high-accuracy measurements in field-deployed scenarios that could never be achieved with classical sensors~\cite{Karcher:2018,Geiger:2020}. Integrated measurement units that fuse high precision, high bandwidth classical sensors with the long-term stability provided by cold-atom inertial sensors, within a portable and rugged form factor, could enable currently unachievable capabilities in navigation, mineral resource extraction and recovery, hydrology and groundwater monitoring, and satellite gravimetry~\cite{Bongs:2019}. Towards this end, there is considerable ongoing research and development into \emph{guided} cold-atom gyroscopes, more suitable for moving platforms~\cite{Jo:2007, Wu:2007, Burke:2009, Qi:2017}, with recent experiments demonstrating per-shot sensitivities of $10^{-5}$ rad/s~\cite{Moan:2020}. 

\subsection{Potential benefits of quantum entanglement}
Equation~(\ref{Delta_a}) shows that there are only four routes to improved cold-atom accelerometer sensitivity: (1) increase the interferometer time $T$, (2) increase the atomic flux, (3) increase the momentum imparted to the atoms by the beamsplitters and mirrors, and (4) surpass the SNL with quantum entanglement ($\xi < 1$). Not all routes are necessarily available. For example, a given size, weight, and power places a fundamental bound on $T$ (through the device size) and also on the maximum $k_\parallel$ (due to both device size and maximum available power). Atomic fluxes are limited by device duty cycles, trapping depths and geometries, and requirements on source coherence and momentum width~\cite{Debs:2011,Szigeti:2012,Robins:2013}. Similar limitations hold for cold-atom gyroscopes [Eq.~(\ref{Delta_Omega})]. Quantum enhancement may therefore be required to achieve the ambitious sensitivity improvements needed for future precision measurements, such as weak equivalence principle tests capable of ruling in or out candidate theories of quantum gravity~\cite{Aguilera:2014,Williams:2016,Becker:2018,Tino:2020}. 

The high accuracy of current cold-atom inertial sensors could also be improved by quantum entanglement. As in the case of atomic clocks, quantum enhancement would allow a cold-atom inertial sensor to reach a given precision faster than a shot-noise-limited inertial sensor. This would accelerate the characterization of systematic errors, potentially increasing the accuracy.

When operating at the environmental noise floor, improvements to sensitivity do not give improved performance. However, a quantum-enhanced sensor could still be beneficial here. For instance, according to Eq.~(\ref{Delta_a}), improvements to sensitivity can be traded for a reduction in $T$, and therefore an increase in measurement rate (bandwidth) and/or a decrease in device size. Concretely, 10 dB of spin squeezing ($\xi = 0.1$) could let you build an accelerometer a factor of 10 smaller for the same sensitivity as a shot-noise-limited device.

\subsection{Requirements for quantum-enhanced cold-atom inertial sensors}
Precision cold-atom inertial sensors require the creation and manipulation of well-defined motional atomic matter-wave modes that are space-time separated. For high fringe contrast, coherence needs to be maintained between these modes for significant periods of interrogation time, alongside good mode-matching at the interferometer output. This must be achieved with large atom number sources and with minimal atom-atom interactions (to minimize phase diffusion~\cite{Altin:2011, Haine:2018}). To realize a stable device that does not drift over long time scales -- a key advantage of cold-atom sensors -- the response of the sensor is locked to some atomic transition. Ideally, the sensor's response should only depend on fundamental constants, rather than the particular macroscopic properties of the device.

State-of-the-art cold-atom inertial sensors meet these requirements, and so must a useful quantum-enhanced cold-atom inertial sensor. However, meeting these requirements whilst also generating entanglement between motional modes and preserving it over large space-time separations is exceedingly difficult. Any experimental imperfection that causes atom loss in the two modes that define the atom interferometer arms will degrade metrologically-useful entanglement within the interferometer, and therefore the degree of quantum enhancement. In current state-of-the-art atom interferometers, such losses are primarily due to detection noise, multimode excitations into modes other than the two interferometer modes (e.g. populating the intermediate momentum states for Bragg pulses~\cite{Altin:2013}) and atoms leaving the laser beam and detection regions (and therefore the interferometer)~\cite{Geiger:2020}. Furthermore, mode-matching requirements are far more stringent for quantum-enhanced atom interferometry. In a shot-noise-limited atom interferometer, imperfect spatial-mode overlap simply degrades the sensitivity via a multiplicative factor in the fringe contrast, whereas in a quantum-enhanced atom interferometer it leads to both a loss of fringe contrast and an increase in fluctuations. The latter effect can be significant; for instance, in the numerical simulations of spin-squeezed Bose-Einstein condensates (BECs) reported in Fig.~3 of Ref.~[\cite{Szigeti:2020}], when imperfect mode matching reduced the average spin length $|\langle \hat{\textbf{J}} \rangle |$ by 35\% in a spherical BEC of $N = 10^6$ atoms, it also reduced the spin squeezing from an anticipated $\xi \approx 0.2$ to zero ($\xi = 1$). Imperfect mode overlap can also manifest as a rotation of the state on the Bloch sphere. For states with a large degree of spin squeezing, this can couple noise from the anti-squeezed spin direction into the measured spin-axis, degrading the sensitivity to significantly worse than the SNL. 

Meeting these stringent requirements has proven challenging, to the degree that a quantum-enhanced measurement of an inertial quantity with a cold-atom sensor has not yet been demonstrated, even in a proof-of-principle laboratory-based device. Indeed, despite sophisticated demonstrations of sub-SNL atom interferometry between internal states, only recently has an experiment converted entanglement between internal states into entanglement between well-separated, controllable motional modes suitable for inertial sensing~\cite{Anders:2020}. An additional challenge is the dearth of precision cold-atom inertial sensors that operate at the SNL. Naturally, quantum-enhancement is only beneficial if a device's noise floor is below the SNL. Although precision shot-noise limited atom interferometry has been demonstrated in a number of clock-like experiments~\cite{Santarelli:1999,McGuirk:2001,Bize:2005}, there has only been one precision cold-atom inertial measurement at the SNL~\cite{Gauguet:2009}. This suggests that a concerted effort to reduce technical noise sources in precision cold-atom inertial sensors may be needed to realize useful quantum-enhanced cold-atom inertial sensing. 

\section{Methods of generating metrologically-useful entanglement}
Quantum metrology with nonclassical atomic states is comprehensively reviewed elsewhere~\cite{Pezze:2018}. Here, we briefly survey both experimentally-realized and proposed entanglement-generating schemes, highlighting their relative strengths and weaknesses for useful quantum-enhanced cold-atom sensing. Broadly, these schemes fall into two classes:
\\\\
\textit{\textbf{Entanglement via atom-light interactions --- }}Entanglement between atomic and photonic degrees of freedom requires strong atom-light coupling, achievable with controllable, high intensity optical fields and atomic ensembles with large optical depth. Optical depth can be substantially increased with an optical cavity, enabling large inter-particle entanglement in high-flux cold-atom thermal sources. Furthermore, since atom-light interactions are central to precision cold-atom sensing, atom-light entangling techniques are largely compatible with current state-of-the-art cold-atom sensing technology. Below we summarize three atom-light entangling approaches: (1) \emph{cavity-mediated coherent atom-light interactions}, (2) \emph{quantum non-demolition (QND) measurements}, and (3) \emph{quantum state transfer and super-radiance}.
\\\\
\textit{\textbf{Entanglement via atom-atom interactions --- }}Ultracold atomic sources, such as BECs, can have high densities and therefore significant collisions between atoms. These atom-atom interactions are highly controllable and a mature experimental approach to generating entanglement. Ideally, the BEC is configured as an effective two-mode system, allowing the atom-atom interactions to generate entanglement between the two interferometer modes. These interferometer modes might be two internal hyperfine atomic states, as used in a clock, or two distinct motional modes, as required for inertial sensing. Below we summarize four atom-atom entangling approaches: (1) \emph{spin-exchange collisions}, (2) \emph{molecular disassociation}, (3) \emph{four-wave mixing}, and (4) \emph{self-phase modulation via atom-atom interactions}. In the first three of these approaches, the exchange of particles directly creates number correlations between two modes, whereas in the fourth the interactions create entanglement between the relative population and relative phase degrees of freedom, which is then converted into number correlations through interference. \\

Both atom-light and atom-atom entangling interactions offer viable approaches to useful quantum-enhanced atomic clocks, cold-atom inertial sensors, and atomic magnetometers.

\subsection{Methods of generating metrologically-useful entanglement for atomic clocks}
The three approaches surveyed below have provided a demonstrated quantum-enhancement to an atomic clock. We therefore consider them to be the `front-running' technologies most likely to result in a useful enhancement to precision timing in the near term.

\subsubsection{Cavity-mediated coherent atom-light interactions} \label{cavity_med}
\begin{figure}[t]
	\begin{center}
				\includegraphics[width=\columnwidth]{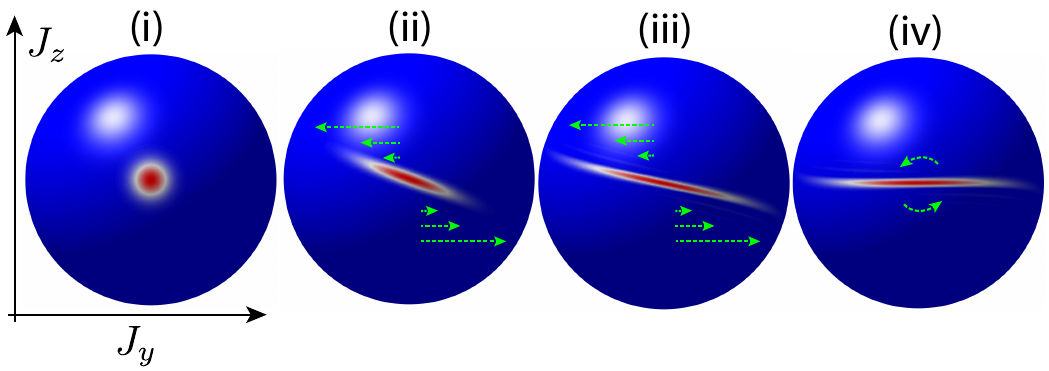}
		\caption{Under the OAT Hamiltonian $\hat{H} = \hbar \chi \hat{J}_z^2$, the Wigner quasiprobability distribution for an initial CCS state (i) is sheared on the Bloch sphere [(ii) and (iii)], creating a spin-squeezed state by reducing the variance of the distribution in one direction at the expense of an increase in the other. To exploit this spin squeezing in the scheme depicted in Fig.~\ref{fig:signal_and_noise}, the state needs to be rotated such that the variance is minimized along $J_z$ (iv).}
		\label{fig:OAT}
	\end{center}
\end{figure}

When atoms are coupled to an optical cavity, light passing through the atoms experiences a refractive-index change that shifts the cavity resonance in proportion to the number of atoms. When the light is highly detuned from the cavity resonance, a positive fluctuation in atom number reduces the detuning from resonance, thereby increasing the steady-state cavity photon number. A negative atom number fluctuation has the opposite effect. This entangles the population degrees of freedom of the atomic and optical fields. Since the light intensity also shifts the internal atomic energies, this atom-light population correlation gets converted into an effective nonlinear self-phase modulation in the atoms, akin to the optical Kerr effect~\cite{Schleier-Smith:2010b}. This effective interaction can be engineered to cause one-axis twisting (OAT)~\cite{Kitagawa:1993}, described by the Hamiltonian $\hat{H} = \hbar \chi \hat{J}_z^2$. Within the Bloch sphere representation of the quantum state, this causes a shearing along the Bloch sphere which, to lowest order, results in an ellipse (Fig.~\ref{fig:OAT}). Consequently, the variance in one direction (minor axis) is decreased at the expense of an increase in a perpendicular direction (major axis), resulting in a spin-squeezed state that can be used to reduce relative number fluctuations at the interferometer output~\cite{Kitagawa:1993}.

The strength and duration of this effective OAT is highly controllable; it can therefore be used to generate spin squeezing in dilute atomic samples and furthermore can be switched-off once the desired spin squeezing is obtained, minimizing unwanted systematic effects due to the atom-light coupling. Cavity-mediated effective OAT has generated 5.6~dB ($\xi \approx 0.275$) of spin squeezing in a cold-thermal ensemble of $^{87}$Rb atoms~\cite{Leroux:2010}. This spin squeezing allowed an atomic clock to reach the same sensitivity 2.8 times faster than a shot-noise-limited atomic clock over 50~s of averaging time~\cite{Leroux:2010b}. Subsequently, cavity-mediated effective OAT has created spin-squeezed states with an 8~dB reduction in variance ($\xi = 0.4$) in larger atom ensembles ($5 \times 10^5)$~\cite{Hosten:2016b}. Most recently, this technique was used in a demonstration of the first squeezed optical lattice clock~\cite{Pedrozo:2020}, achieving $\xi = 0.60$.

Commonly, in a cavity-mediated atom-light interaction (coherent or otherwise), the participating atoms are not addressed identically, especially during the generation of the squeezing and the readout. In this case, the definition of the spin-squeezing parameter Eq.~(\ref{Delta_phi_spinsqz}) still holds, with a collective spin operator of reduced length $|\langle \bm{\hat{J}}\rangle|<N/2$, provided the individual addressing strengths remain constant~\cite{Hu:2015}. Hence, although nonuniformities pose little threat for applications like optical lattice clocks where atoms can be locked in position, the uniform generation and readout of squeezed states is a critical requirement for fountain clock applications where atoms need to be in free motion~\cite{Wu:2020,Malia:2020}, as well as for inertial sensing.

\subsubsection{Quantum non-demolition (QND) measurements} \label{QND_meas}
\begin{figure}[h]
	\begin{center}
	\includegraphics[width=\columnwidth]{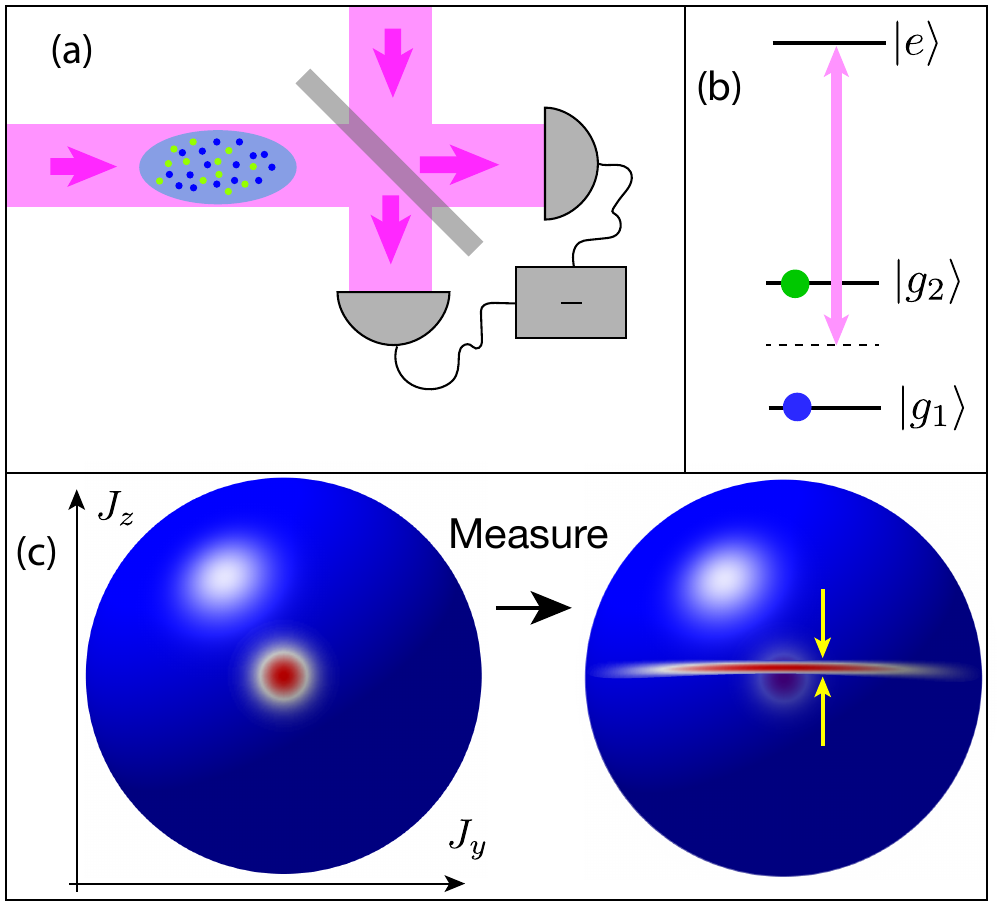}
		\caption{Schematic of spin squeezing via a QND measurement. (a) Atoms with two hyperfine groundstates $|g_1\rangle$ and $|g_2\rangle$ and an excited state $|e\rangle$ are illuminated with light, which is subsequently interfered with a local oscillator in order to perform a homodyne measurement. (b) In order to engineer a dispersive probe where the photon flux after atom-light interactions is proportional to $\hat{J}_z$, the frequency of the probe laser is chosen such that the detuning of $|g_1\rangle \to |e\rangle$ and $|g_2\rangle \to |e\rangle$ is $\pm \Delta$, respectively. (c) Performing a QND measurement on a CSS with spin projection $J_x = N/2$ creates a spin-squeezed state with reduced variance in $\hat{J}_z$, shifted off the equator of the Bloch sphere by an amount conditional on the homodyne measurement result.}
		\label{fig:QND}
	\end{center}
\end{figure}

Dispersive probing of an atomic ensemble can give indirect information about an atomic observable such as $\hat{J}_z$. This information allows an experimenter to be more certain of the relative number difference, thereby reducing the variance in $\hat{J}_z$ below $N/4$ and giving a spin-squeezed state~\cite{Kuzmich:1998} (see Fig.~\ref{fig:QND}). QND measurements are indirect in the sense that the atom-light coupling does not change the value of the measured observable ($\hat{J}_z$) - i.e. the atom-light coupling Hamiltonian commutes with the observable. 

Many early atomic spin-squeezing experiments were accomplished with QND measurements of cold-thermal ensembles in free space~\cite{Appel:2009,Takano:2009,Louchet_Chauvet:2010,Koschorreck:2010,Sewell:2012}. These experiments demonstrated only modest amounts of spin squeezing (1-3~dB) in ensembles of $10^5-10^6$ atoms. However, since the degree of spin squeezing (in the linear spin-squeezing regime) is directly proportional to the atom-light coupling strength, cavity-assisted QND measurements have generated substantially larger degrees of spin squeezing~\cite{Schleier-Smith:2010b}, including a record 20~dB suppression of spin variance in $10^6$ atoms~\cite{Hosten:2016}. This latter experiment used this QND-spin-squeezed state to achieve a microwave Rb atomic clock measurement a factor of 11 below the SNL. Cavity-assisted QND measurements have also recently provided spin squeezing for an atomic fountain clock operating in free space~\cite{Malia:2020}. Although QND measurements are typically associated with off-atomic-resonance (dispersive) probes, measurements to the same effect in the near-atomic-resonance regime have also been demonstrated by probing the vacuum Rabi splitting in a cavity~\cite{Chen:2011}.

\subsubsection{Spin-exchange collisions} \label{spin_exchange_collisions}
\begin{figure}[h]
	\begin{center}
		\includegraphics[width=\columnwidth]{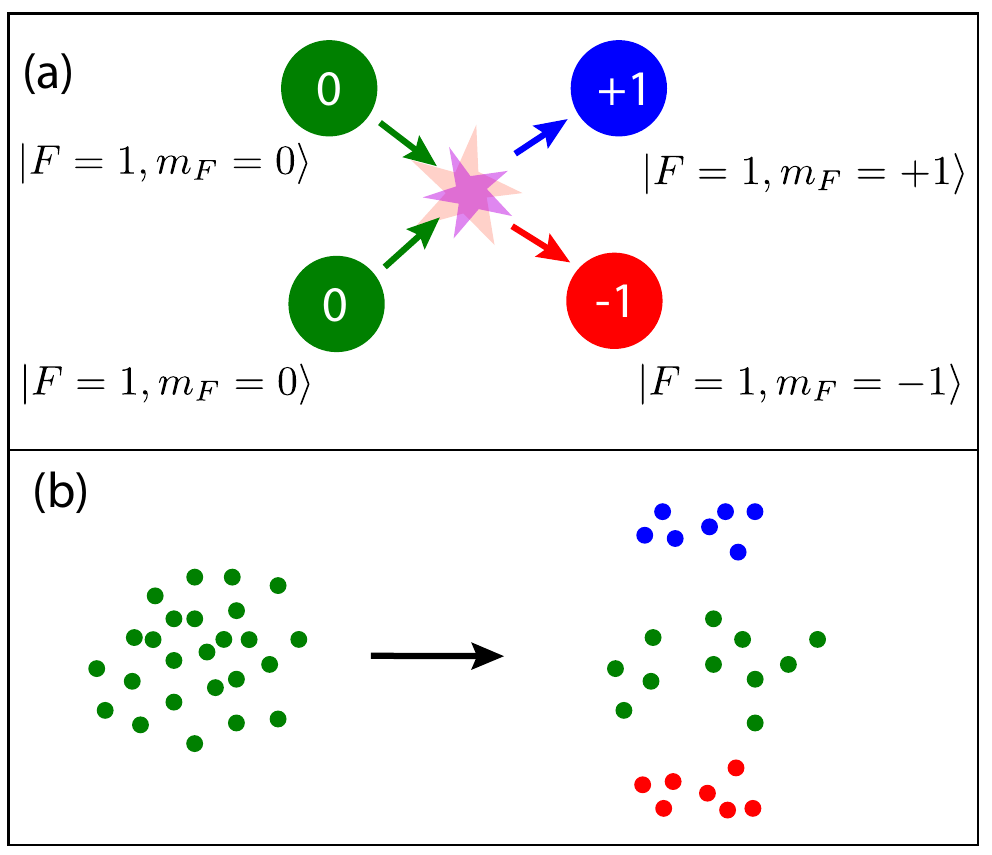}
		\caption{(a) Illustration of a spin-changing collisional process in a spin-1 gas, where two atoms in internal state $|F = 1, m_F = 0\rangle \equiv |0\rangle$ collide, resulting in two scattered atoms in states $|F = 1, m_F = \pm 1\rangle \equiv |\pm 1\rangle$. (b) Over time, some of the atoms in $|0\rangle$ (green) are converted into correlated (entangled) pairs of atoms in $|+1\rangle$ (blue) and $|-1\rangle$ (red).}
		\label{fig:spin_exchange}
	\end{center}
\end{figure}

\begin{figure}[h]
	\begin{center}
		\includegraphics[width=\columnwidth]{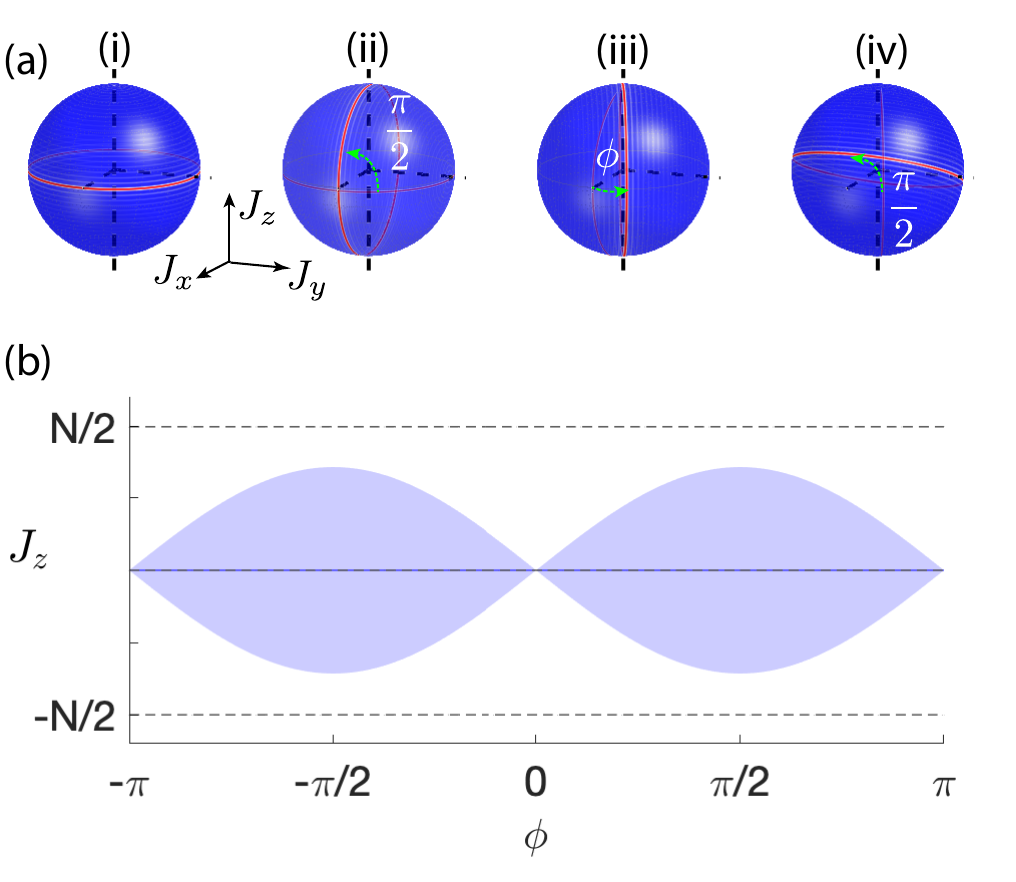}
		\caption{(a) Evolution of a twin-Fock state's Wigner quasiprobability distribution (i) after a $\pi/2$ pulse (ii), a relative phase shift $\phi$ between two interferometer modes (iii), and a final $\pi/2$ pulse (iv). This is the same interferometer sequence depicted in Fig.~\ref{fig:signal_and_noise}(a). (b) Performing a $\hat{J}_z$ measurement on the output state (a, iv) gives mean signal $\langle \hat{J}_z \rangle = 0$ for all values of $\phi$. However, the fluctuations $\langle \hat{J}_z^2 \rangle$ (indicated by the shaded region) are clearly $\phi$-dependent. As $\phi \to 0$, the variance $\text{Var}(\hat{J}_z^2)$ approaches zero faster than $\partial \langle \hat{J}_z^2 \rangle / \partial \phi$, yielding a finite value of $\Delta \phi = \sqrt{\text{Var}(\hat{J}_z^2)} / |\partial \langle \hat{J}_z^2 \rangle / \partial \phi| \to \sqrt{2} / N$ for $N \gg 1$~\cite{Kim:1998,Haine:2015b}.}
		\label{fig:twinfock}
	\end{center}
\end{figure}

In spinor BECs, colliding atoms can exchange spin~\cite{Stamper-Kurn:2013}. Consider, for instance, a spin-1 gas where the three spin projections along some quantization axis define the three-level system: $|0\rangle \equiv |F=1, m_F=0\rangle$ and $|\pm 1\rangle \equiv |F=1, m_F=\pm1\rangle$. Provided an energy-resonance condition is satisfied (e.g., by adjusting each state's energy through the quadratic Zeeman shift), a collision between two $|0\rangle$ atoms can exchange spin, resulting in a $|+1\rangle$ atom and a $|-1\rangle$ atom~\cite{Schmaljohann:2004, Chang:2005} (see Fig.~\ref{fig:spin_exchange}). Conservation of angular momentum ensures that the number of $|+1\rangle$ and $|-1\rangle$ atoms produced is identical, yielding sub-Poissonian number statistics~\cite{Lucke:2011, Gross:2011,Bookjans:2011, Hamley:2012, Zou:2018}. These number correlations are associated with particle entanglement and allow relative-phase measurements below the SNL. As with all schemes that add correlated pairs of particles to vacuum, the entangled state generated within the $\{|-1\rangle,|+1\rangle\}$ subspace is not a spin-squeezed state, since $\langle \hat{\textbf{J}}\cdot \textbf{n}\rangle = 0$ for every unit vector $\textbf{n}$ and so the Wineland spin-squeezing parameter is undefined. Consequently, a relative phase shift $\phi$ between $|-1\rangle$ and $|+1\rangle$ cannot be inferred from a $\hat{J}_z$ measurement (see Fig.~\ref{fig:twinfock}). However, since fluctuations in the spin depend on $\phi$, sub-SNL sensitivities are still achievable by choosing $\hat{S} = \hat{J}_z^2$. Unlike standard spin-squeezed metrology, where the best sensitivity occurs when the derivative of the signal is maximal, here the sensitivity is best when both the numerator and denominator of Eq.~(\ref{Delta_phi}) approach zero, with the numerator approaching zero faster than the denominator such that the limit approaches the Heisenberg limit $\Delta \phi = 1/N_s$. Here $N_s$ is the number of atoms outcoupled from $|0 \rangle$ into the $\{|-1\rangle,|+1\rangle\}$ subspace via spin-exchange collisions, and is necessarily smaller than the total number of atoms $N$ initially prepared in $|0 \rangle$. Without single-atom resolving detection, the enhancement is significantly worse than this Heisenberg limit, as the additional noise affects the numerator of Eq.~(\ref{Delta_phi}) much more strongly than it does the denominator. For instance, Ref.~[\cite{Lucke:2011}] reported $\xi \approx 0.83$ with $N = 8,000$ atoms, limited predominantly by detection noise of $\pm$ 20 atoms. This fragility to inefficient detection is a common feature of metrology with correlated particle pairs~\cite{Demkowicz-Dobrzanski:2012, Gabbrielli:2015, Macri:2016}.

Quantum-enhancement with spin-exchange collisions is not restricted to the $\{|-1\rangle,|+1\rangle\}$ subspace. For example, interfering correlated pairs in $|\pm 1\rangle$ with atoms in $|0\rangle$ allowed a sub-SNL measurement of the $^{87}$Rb atomic clock transition with $\xi = 0.62$~\cite{Kruse:2016}. Although spin-exchange collisions only created a small number of correlated pairs ($N_s = 0.75$ atoms on average), coherently mixing these pairs with the remaining atoms in $|0\rangle$ meant that the measurement was made with all $N = $10,000 atoms in the ensemble, giving $\Delta \phi = \xi / \sqrt{N}$. This is both robust to imperfect detection and significantly better than the Heisenberg limit $1/N_s$. Pumped-up SU(1,1) interferometry operates on a similar principle~\cite{Szigeti:2017}.

Spin systems are sensitive to magnetic fields~\cite{Vengalattore:2007, Sewell:2012}, so quantum-enhanced magnetometers could exploit the entanglement created by spin-exchange collisions~\cite{Sompet:2019}. Conversely, magnetic sensitivity makes these systems less appealing candidates for inertial sensors. Nevertheless, with adequate magnetic shielding it may be possible to convert entanglement via spin-exchange collisions into entangled momentum states suitable for inertial sensing, as discussed in Sec.~\ref{Entanglement_Methods_Particle_Exchange}.

\subsection{Methods of generating metrologically-useful entanglement for cold-atom inertial sensing}
Unlike atomic clocks, there has not yet been a demonstration of a quantum-enhanced cold-atom inertial sensor. Nevertheless, there are a variety of approaches that generate the needed entanglement between atomic momentum modes, or at least generate entanglement between internal atomic states that could feasibly be converted to momentum-state entanglement.

\subsubsection{Cavity-assisted atom-light interactions}
As surveyed in Sec.~\ref{cavity_med} and Sec.~\ref{QND_meas}, cavity-mediated OAT and cavity-enhanced QND measurements have successfully generated large degrees of spin squeezing in atomic ensembles. However, this entanglement is between the internal states of the atoms; to date such approaches have not yet provided the entangled momentum states needed for quantum-enhanced inertial sensing. Nevertheless, many recent proposals for quantum-enhanced cold-atom inertial sensing exploit QND measurement~\cite{Salvi:2018, Shankar:2019, Kritsotakis:2021}. Furthermore, the recent demonstration of a QND-squeezed atomic fountain clock in free space~\cite{Malia:2020} shows that entanglement between internal atomic states persists as atomic clouds spatially expand, a necessary (but not sufficient) requirement for converting such internal-state entanglement into momentum-mode entanglement suitable for precision cold-atom inertial sensing.

\subsubsection{Quantum state transfer and super-radiance}
The successful enhancement of gravitational wave detection via entangled (squeezed) light sources speaks to the maturity of squeezed-light-generation technology. Squeezed light therefore offers a controllable means of creating entangled atomic states by transferring the quantum state of the squeezed light onto the atomic field~\cite{Moore:1999,Jing:2000,Fleischhauer:2002b, Haine:2005, Haine:2005b, Haine:2006b, Hammerer:2010}. This requires an atom-light interaction, such as a Raman transition, that correlates the absorption and emission of a photon with the transfer of an atom into a specific internal state and/or momentum mode. Furthermore, since the squeezed light is generated independently to the atomic source, this approach leverages existing cold-atom and squeezed-light technology, making it highly compatible with the requirements of current high-precision cold-atom sensors. Nevertheless, an atomic spin-squeezed state created via the quantum state transfer of squeezed light has never been used in a quantum-enhanced atom interferometer. One major challenge is achieving a sufficiently high efficiency quantum state transfer process. Atomic spin squeezing via the quantum state transfer of squeezed light has been experimentally demonstrated~\cite{Hald:1999}. However, generating the narrow momentum width atomic wavepackets needed for high precision atom interferometry requires high efficiency quantum state transfer in a very narrow band of the optical squeezing, which has yet to be demonstrated. This can be somewhat overcome by \emph{information recycling}, whereby the portion of the photonic field not mapped onto the atomic field is measured and this information appropriately combined with the atom interferometer measurement signal ($\hat{S}$)~\cite{Haine:2013,Szigeti:2014b,Tonekaboni:2015,Haine:2015, Haine:2015b}. However, this technique requires complex optical detection in order to extract the information from a narrow frequency band of the light~\cite{Haine:2016}.

A related approach exploits the phenomenon of super-radiance, where Bose enhancement of the emitted photons from an optically pumped atomic sample gives a highly directional emission pattern~\cite{Schneble:2004, Yoshikawa:2004, Cola:2004, Uys:2007, Hilliard:2008}. Each spontaneously emitted photon is correlated with the transfer of one atom into a distinguishable internal state or momentum mode~\cite{Moore:2000}. It was shown that performing a homodyne measurement on these super-radiant photons projects the atomic state into an entangled state that can be used for quantum-enhanced atom interferometery~\cite{Haine:2013, Haine:2016}. Using a coherent Raman transition to create a small coherent seed of atoms forces the super-radiant emission into a particular momentum mode, which can be selected to suit the particular atom interferometric application. Like quantum state transfer, this scheme utilizes the same lasers as the atom interferometer beamsplitting pulses. Consequently, the entangled matter-waves are automatically mode-matched to the atomic beamsplitters, and are therefore suitable as the input state for high-precision atom interferometry. This makes these schemes appealing candidates for quantum-enhanced atom interferometry. Super-radiance techniques may also allow for continuous readout of the gravitational field~\cite{Gietka:2019}.

\subsubsection{Particle-exchange dynamics due to atom-atom collisions} \label{Entanglement_Methods_Particle_Exchange}
For strongly-interacting atomic ensembles, such as BECs, atom-atom interactions can cause particle-exchange dynamics that directly create number correlations between two modes. Below we survey three approaches to generating entanglement with particle-exchange dynamics.
\\\\
\textit{\textbf{a. Spin-exchange collisions ---}} As surveyed in Sec.~\ref{spin_exchange_collisions}, spin-exchange collisions can generate entangled atom pairs suitable for quantum-enhanced atom interferometry. However, generating $\sim 10^4$ entangled atomic pairs (likely a bare minimum for precision inertial sensing with only two modes) requires a high-density atomic ensemble undergoing spin-changing collisions for time scales between hundreds of milliseconds to seconds~\cite{Peise:2015,Luo:2017}. The atoms must therefore be trapped for the duration of the entanglement-generation protocol, resulting in entanglement between two internal atomic states with the same average momentum. A recent experiment has shown that this entanglement between internal states can be converted into entanglement between well-separated momentum states in free space~\cite{Anders:2020}. Here, after preparing a twin-Fock state of $~10^4$ $^{87}$Rb atoms in $|F=1,m_F=\pm 1\rangle$, the atoms are released from the trap and collimated. Subsequently, a microwave pulse transfers atoms from $|F=1,m_F=- 1\rangle$ to $|F=2,m_F=0\rangle$, followed by a Raman pulse that transfers atoms in $|F=2,m_F=0\rangle$ to $|F=1,m_F=0\rangle$, accompanied by a $2 \hbar k$ momentum kick [$k = 2\pi / (780~\textrm{nm}$)]. Although the presence of entanglement between these two momentum modes was successfully verified, these entangled momentum modes were not used to perform a sub-SNL phase measurement. Nevertheless, this demonstration cements spin-exchange collisions as the leading particle-exchange approach to generating entanglement suitable for quantum-enhanced cold-atom inertial sensing.

A similar scheme could be used for quantum-enhanced rotation sensing with a toroidally-trapped spin-1 BEC~\cite{Nolan:2016}. A Raman transition, implemented via Laguerre-Gaussian optical beams, first coherently outcouples small `seeds' of atoms into $|F=1, m_F = \pm 1\rangle$ with orbital angular momentum $\pm \hbar l$, respectively. Bose enhancement then causes a subsequent period of spin-exchange collisional dynamics to preferentially scatter pairs of atoms into these well-defined orbital angular momentum modes, creating a counter-rotating, spin-squeezed state suitable for rotation sensing. As shown in Ref.~[\cite{Kruse:2016}], a large number of entangled atom pairs are not necessarily required for sub-SNL atom interferometry, since these can be coherently combined with the remaining $|F=1, m_F =0\rangle$ atoms to boost the overall atom number in a three-mode interferometer.
\\\\
\textit{\textbf{b. Molecular disassociation ---}} The formation and disassociation of weakly-bound ultracold dimers can be controlled via a Feschbach resonance~\cite{Durr:2004} or photoassociation~\cite{McKenzie:2002}. This could provide correlated atomic pairs which, for sufficiently slow disassociation, would be outcoupled from the molecular BEC as well-defined atomic beams~\cite{Kheruntsyan:2005}. Unfortunately, the momentum of these beams is not naturally mode-matched to an optical beamsplitting process, so using this as a correlated atomic source for quantum-enhanced interferometry would be challenging.
\\\\
\textit{\textbf{c. Four-wave mixing ---}} Atomic four-wave mixing~\cite{Deng:1999} involves the collision and scattering of two incoming matter-waves into two distinct momentum modes. This process is either spontaneous or stimulated and, due to momentum conservation, gives correlated pairs in the two outgoing modes. Consequently, atomic four-wave mixing can create entangled matter-wave momentum modes~\cite{Vogels:2002, Vogels:2003, Zin:2005, Deuar:2007, Ferris:2009, Ogren:2009, Dall:2009, Dennis:2010, Haine:2011}; this has been verified experimentally~\cite{Perrin:2007, Jaskula:2010, Bucker:2011}. For confined BECs, this scattering is not necessarily into the untrapped atomic momentum modes that form the arms of a free-space atom interferometer. For instance, in Ref.~[\cite{Bucker:2011}] four-wave mixing scattered atoms from the ground state of the trapped condensate to the first excited state of the trap.

Unlike spin-exchange collisions, four-wave mixing can occur in magnetically-insensitive states, which is beneficial for inertial sensing. However, as with any method that produces correlated pairs rather than spin-squeezed states, the phase must be inferred from changes in the final state's fluctuations ($\hat{S} = \hat{J}_z^2$), rather than changes in the mean ($\hat{S} = \hat{J}_z$). This makes the sensitivity extremely susceptible to imperfect detection. However, the use of a coherent seed in the target modes to create a preferred direction for the atomic spin alleviates this problem \cite{Haine:2011}.

Four-wave mixing between two distinct internal atomic states of different momentum, $|1,\textbf{k}\rangle$ and $|2, -\textbf{k} \rangle$, can give scattered momentum modes $|2,\textbf{k}\rangle$ and $|1,-\textbf{k}\rangle$ that are co-propagating with the incident momentum modes, but with different internal states~\cite{Pertot:2010}. By appropriately interfering the incident and scattered modes, this four-mode system can be configured as two correlated atom interferometers, whose output signals can be combined to give a sub-SNL phase measurement~\cite{Haine:2011}. This proposal needs to be modified to measure gravity or gravity gradients; more problematically, it suffers from poor mode-matching due to density-dependent depletion from the incident modes.

\subsubsection{Self-phase modulation via atom-atom interactions} \label{self_phase_atom_atom}
Elastic atom-atom collisions where the atoms do not change state can also give significant metrologically-useful entanglement. Even though the atomic interactions do not directly affect the number statistics, the nonlinear-dependence of the energy on the population in each mode causes entanglement between the relative number and relative phase degrees of freedom. This number-phase entanglement can be converted into relative-number ($\hat{J}_z$) squeezing by carefully interfering these modes such that the probability amplitudes corresponding to large number-difference fluctuations destructively interfere. 

The archetypical model of an atomic self-phase modulation induced by atom-atom collisions is OAT (Fig.~\ref{fig:OAT}). Since the interatomic collisions within a BEC are very strong (relative to lower density cold but still thermal sources), OAT has generated significant spin squeezing between two internal states in a BEC, both when the two components are overlapping in a lattice~\cite{Esteve:2008} and also within a double well produced via a chip trap~\cite{Riedel:2010}. However, strong atom-atom collisions are undesirable for precision measurement. In trapped systems, interatomic collisions can generate unwanted multimode dynamics that hinder mode-matching at the interferometer output, degrading both fringe contrast and entanglement~\cite{Haine:2014, Nolan:2018}. Furthermore, atomic interactions couple relative-number fluctuations into phase fluctuations which, without a detailed knowledge of the underlying process, cannot be converted into useful spin squeezing and therefore degrades sensitivity via phase diffusion~\cite{Altin:2011, Haine:2018}. Both effects have limited experimental demonstrations of quantum-enhanced atom interferometry via self-phase modulation to small atom numbers and short interrogation times. In principle, strong atom-atom interactions in a trapped system could be eliminated or reduced via a Feshbach resonance~\cite{Cornish:2000, Fattori:2008, McDonald:2014, Everitt:2017}. However, exploiting a Feshbach resonance requires the detailed control of magnetic fields. Achieving this required control over the full spatial extent of the interferometer, for the duration of its interrogation time, would be very challenging for any precision device. More significantly, unknown fluctuations in the magnetic field can introduce systematic effects that impact the accuracy and long-term stability of the cold-atom sensor.

A recent proposal showed that in a freely-expanding condensate, strong atom-atom interactions can generate substantial squeezing between two momentum modes without degrading mode overlap or causing significant phase diffusion~\cite{Szigeti:2020}. Under free expansion, the atomic interactions that initially cause spin squeezing rapidly `switch off', largely reducing their unwanted effects during most of the interferometer sequence. This scheme is highly compatible with the requirements of high precision cold-atom gravimetry and current cold-atom technology.

There are more sophisticated procedures than OAT for generating spin squeezing via a self-phase modulation with atom-atom interactions. For instance, a radiation pulse can linearly couple the two interferometer modes during atom-atom collisional dynamics, yielding the so-called \emph{twist-and-turn} Hamiltonian~\cite{Muessel:2015,Sorelli:2019}: $\hat{H} = \hbar \chi \hat{J}_z^2 - \hbar \Omega \hat{J}_x$. For an appropriate choice of coupling strength $\Omega$, this can generate spin squeezing at a faster rate than OAT.

\subsection{Methods of generating metrologically-useful entanglement for atomic magnetometers}
Finally, although this Perspective has focussed on atomic clocks and inertial sensors, we briefly summarize experimental demonstrations of quantum-enhanced atomic magnetometers. Atomic magnetometers can provide alternative capabilities to other state-of-the-art magnetic sensing platforms such as superconducting quantum interference devices~\cite{Fagaly:2006} (SQuIDs) and nitrogen vacancy (NV) centres in diamonds \cite{Barry:2020, Mitchell:2020}. In particular, atomic magnetometers provide high sensitivities at low frequencies without the need for cryogenic cooling~\cite{Tierney:2019}. This has provided new capabilities in biological and medical sensing, such as noninvasive detection of nerve pulses~\cite{Jensen:2016}, magnetocardiography~\cite{Jensen:2018}, and the monitoring of neural activity within the brain~\cite{Zhang:2020}.

Atomic magnetometers operate by first preparing a spin-polarized state perpendicular to the component of the field to be measured, frequently but not always by optical pumping, and then observing the Larmor precession of the spin due to the interaction between the atom's spin-dependent magnetic moment and the magnetic field. This can alternatively be viewed as the creation of a coherent superposition of spin components parallel to the magnetic field, each of which develops a phase shift proportional to the strength of the field and the spin projection. That is, atomic magnetometry operates via the principles of atom interferometry, and therefore is subject to the same atomic shot-noise limits.

The first demonstrations of quantum-enhanced atomic magnetometry were performed in optically-pumped hot atomic vapours, where spin squeezing was generated by using the Faraday rotation of light to perform a QND measurement (see Sec.~\ref{QND_meas}). With a sample of approximately $1.5\times 10^{12}$ caesium atoms, Wasilewski \emph{et al.} measured a weak AC magnetic field (oscillating at the Larmor frequency) at a sensitivity 1.5~dB below the atomic SNL, and at an absolute sensitivity in the femtoTesla/$\sqrt{\mathrm{Hz}}$ regime~\cite{Wasilewski:2010}. This is comparable to, but not surpassing, state-of-the-art sensitivities achieved by optically pumped atomic magnetometers in much larger ensembles \cite{Lee:2006, Dang:2010}. In principle, this approach to quantum-enhanced atomic magnetometry should be viable for larger numbers of atoms, and so could potentially improve upon current magnetic sensing capabilities. A similar QND-squeezing approach has improved the bandwidth of an atomic magnetometry measurement~\cite{Shah:2010}. Entanglement-enhanced magnetic sensing close to state-of-the-art has also been demonstrated using cold, dipole-trapped ensembles of atoms~\cite{Koschorreck:2010, Sewell:2012}.

Bose-condensed atoms have also been used to perform quantum-enhanced magnetometry~\cite{Ockeloen:2013, Muessel:2014}. In both experiments, atom-atom collisions caused OAT dynamics (see Sec.~\ref{self_phase_atom_atom}), generating a spin-squeezed state that was subsequently transferred to magnetically-sensitive atomic levels. Ockeloen \emph{et al.} demonstrated 4~dB of magnetically-sensitive spin squeezing in a sample of 1400 atoms~\cite{Ockeloen:2013}, whereas Muessel \emph{et al.} demonstrated 5.5~dB with 12,300 atoms\cite{Muessel:2014}. Due to the small volume of Bose-condensed samples, relative to thermal vapours, these experiments were close to state-of-the-art in terms of sensitivity per unit volume~\cite{Mitchell:2020}. However, the duty cycle of these experiments was limited by the $\gtrsim 10$~ms of state-preparation time, compared to less than a millisecond of interrogation time. Still, such quantum-enhanced sensors may enable niche capabilities that require high spatial resolution of the magnetic field.

\subsection{Entanglement Beyond the Gaussian Regime}
\begin{figure}[t]
	\begin{center}
		\includegraphics[width=0.5\columnwidth]{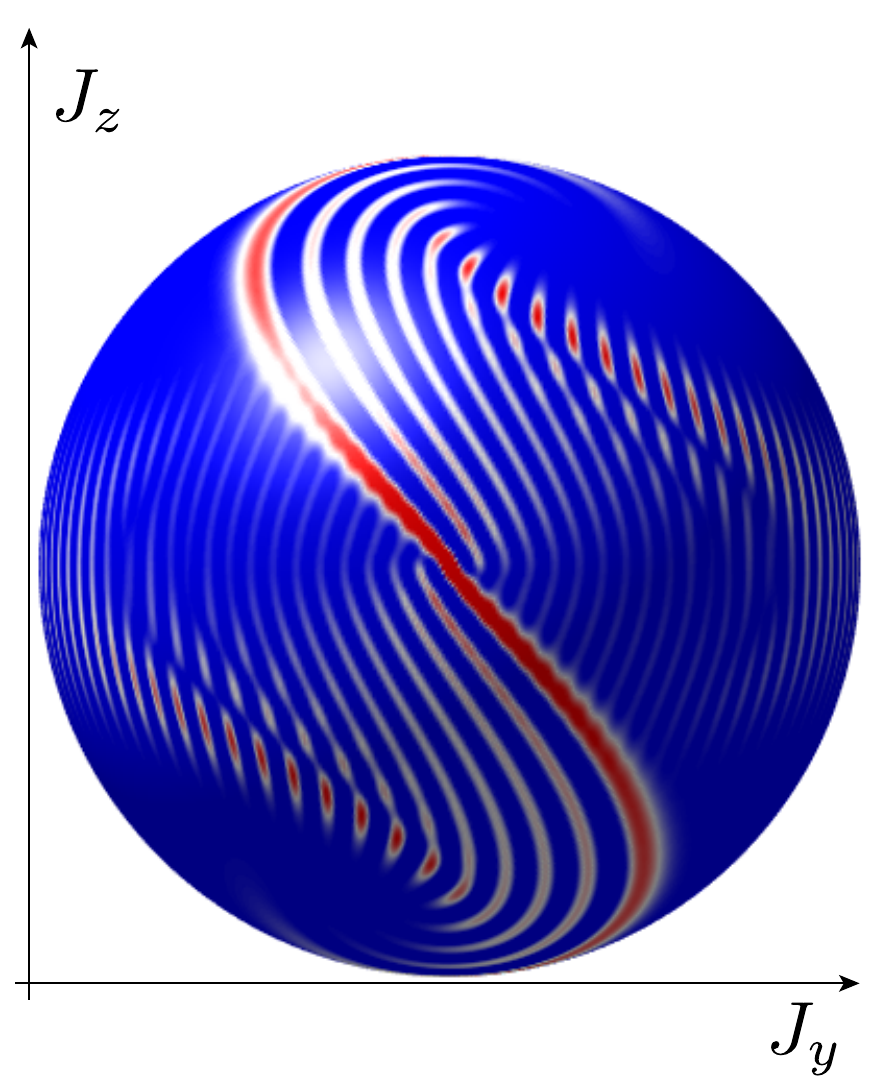}
		\caption{The Wigner quasiprobability distribution of an entangled non-Gaussian state that does not display spin squeezing, yet has abundant metrologically-useful entanglement. Exploiting this metrological potential for a useful measurement of a physical quantity requires a much more difficult readout scheme than that depicted in Fig.~\ref{fig:signal_and_noise}.}
		\label{fig:ENGS}
	\end{center}
\end{figure}

As surveyed above, all experimentally-demonstrated quantum-enhanced atom interferometers with large (i.e. useful) atom numbers have exploited Gaussian spin-squeezed states. However, in principle more exotic non-Gaussian entangled states exist that have superior metrological potential~\cite{Pezze:2018} (see Fig.~\ref{fig:ENGS}). For example, for a sufficiently long interaction time, OAT results in a non-Gaussian state that allows sensitivities approaching the Heisenberg limit $\Delta \phi = 1/N$~\cite{Agarwal:1997,Pezze:2009}. Experimentally characterizing highly-entangled non-Gaussian states cannot be done with the Wineland spin-squeezing parameter, since it requires the measurement of higher-order moments of the collective spin operators. This has prompted several generalizations of the spin-squeezing parameter suitable for states that are not spin squeezed~\cite{Hyllus:2012b,Lucke:2014,Vitagliano:2014,Gessner:2019}. Proof-of-principle experiments have observed non-Gaussian entangled states in modest atom number ensembles~\cite{Lucke:2011,Strobel:2014,Haas:2014,McConnell:2015,Luo:2017} and even realized Heisenberg-limited phase sensitivities (strictly $\Delta \phi = 1 / \sqrt{N_s(N_s+2)}$) in an atomic SU(1,1) interferometer~\cite{Linnemann:2016} (albeit with only $N_s = 2.8$ atoms on average). Generating such non-Gaussian entangled states in larger atom ensembles and incorporating them into high precision cold-atom sensors is a daunting prospect. These states are exceedingly fragile, and therefore place more stringent requirements on mode-matching, measurement resolution, and tolerance to losses than spin-squeezed states~\cite{Pezze:2018, Brif:2020, Nolan:2017}. Methods of alleviating these requirements and improving robustness by subsequently undoing the entanglement generation after the phase-encoding (so called `interaction-based readouts') is an active field of research~\cite{Hosten:2016b, Davis:2016, Davis:2017, Frowis:2016, Macri:2016, Linnemann:2016, Nolan:2017b, Mirkhalaf:2018, Haine:2018b,Anders:2018,Haine:2020, Fang:2020, Schulte:2020}.

Despite their fragility, under the right circumstances entangled non-Gaussian states could prove beneficial even in presence of noise. An example is when a protocol is limited not by the atomic decoherence time, but rather by the coherence time of the probing laser. This is the case for current state-of-the-art atomic clocks. Groups of sequentially larger GHZ states can improve clock stabilities by operating near the Heisenberg limit~\cite{Kessler:2014}. In comparison to an unentangled clock, the cascaded GHZ scheme provides $\sqrt{N/\log N }$ stability enhancement for short averaging times and reaches the fundamental noise floor afforded by the presence of single particle decoherence $N/\log N $ times faster. Recently, GHZ states with record numbers of particles were demonstrated in trapped ion systems (24 particles)~\cite{Pogorelov:2021} and in Rydberg atom arrays (20 particles)~\cite{Omran:2019}. These platforms are prime candidates for exploring the use of GHZ states in clocks based on trapped ions~\cite{Brewer:2019} and tweezer arrays~\cite{Young:2020}.

All these nascent techniques certainly warrant further investigation. However, in the near-to-medium term, it is unlikely that entangled non-Gaussian states will give practical improvements to precision cold-atom sensing.

\section{Outlook for quantum-enhanced cold-atom sensing}
The challenge in developing a useful quantum-enhanced cold-atom sensor is not the generation of entanglement \emph{per se}, but rather generating entanglement in a way that is compatible with the requirements of a given precision sensing application. For instance, 10~dB of spin squeezing cannot result in a useful improvement if it can only be achieved in small atom numbers, for small interrogation times, and in the absence of common experimental noise sources. Future research needs to focus on using entanglement to improve \emph{absolute} sensitivity over simply surpassing the SNL, and doing so in a manner that does not compromise one of the key advantages of cold-atom sensors over classical sensors: their long-term stability. 

Prospects for useful quantum-enhanced atomic clocks are extremely promising. Entanglement has been used to enhance the short-term sensitivity of both microwave~\cite{Leroux:2010b,Hosten:2016,Malia:2020} and optical clocks~\cite{Pedrozo:2020} (up to 10 dB and 4.4 dB respectively), improving the averaging times needed to reach a given sensitivity. The survival of the entanglement itself at second-timescales has also been demonstrated~\cite{Huang:2020}. QND measurements and cavity-mediated OAT stand out as the main methods of achieving the entanglement. Within the next decade it seems reasonable to expect portable atomic clocks running with 10~dB entanglement enhancement. For high end laboratory clocks, given realistic noise analyses~\cite{Schulte2:2020}, an absolute quantum entanglement supremacy might be expected for the case of optical lattice clocks where employable particle numbers are relatively limited.

The path towards useful quantum-enhanced cold-atom inertial sensing is a longer one, and therefore more uncertain. An essential first step is a proof-of-concept sub-SNL measurement of an inertial quantity such as acceleration or rotation rate. Developments in quantum-enhanced clocks definitely provide lessons that inform design strategies for this goal. Proposals that realize the needed entangled momentum states via cavity-enhanced QND measurements are particularly well-developed~\cite{Salvi:2018, Shankar:2019} and, given the demonstrated success of spin squeezing via QND measurements~\cite{Malia:2020,Wu:2020}, are strong candidates for proof-of-principle demonstrations. Indeed, the strength of the atom-light interaction within a cavity, and its controllability, also suggests cavity-mediated OAT~\cite{Schleier-Smith:2010b, Leroux:2010} as a viable approach, although as with QND measurements, the key challenge is generating entanglement between well-defined momentum modes, rather than atomic internal states. Appropriately configured super-radiance protocols~\cite{Haine:2013, Gietka:2019} circumvent this by directly entangling light with atomic motional modes. However, all of these schemes require an optical cavity. Optical cavities impart additional advantages to atom interferometers, allowing higher efficiency large momentum transfer beamsplitting for the same power, favourable mode filtering, and improved detection resolution~\cite{Poldy:2008}. On the other hand, using optical cavities for entanglement generation and/or atom interferometry increases experimental complexity and SWaP, which may be unsuitable for inertial sensing applications that need compact, low-weight, low-power, and portable devices. Quantum state transfer directly entangles the appropriate momentum modes without the need for an optical cavity~\cite{Szigeti:2014b}, but also suffers from additional experimental overhead and complicated optical detection, and has yet to be demonstrated with sufficient transfer efficiency. Inertial sensing with a BEC provides an attractive alternative to cavity-based approaches here, since the atom-atom interactions can be used to generate entangled momentum modes in free space without the need for optical guiding or cavities. In particular, spin-exchange collisions can generate entanglement between internal atomic states, which can be transferred to well-separated momentum modes suitable for inertial sensing~\cite{Anders:2018}. Furthermore, a BEC's atom-atom interactions can induce a self-phase modulation that produces spin-squeezed momentum modes in free space~\cite{Szigeti:2020}. Indeed, this latter proposal requires only a small modification to the pulse sequences of current state-of-the-art laboratory setups, also making it a promising candidate for a near-term proof-of-concept quantum-enhanced cold-atom inertial sensor.

It is unlikely that a single approach will be suitable for all sensing applications. For instance, clocks have different requirements than inertial sensors, and the requirements of a highly controlled laboratory-based experiment differ greatly to a portable field-deployed device with low SWaP and robustness to vehicle motion. Fortunately, cold-atom systems are highly controllable and can be flexibly configured which, as surveyed above, has already allowed a myriad of demonstrated approaches to entanglement generation in cold atoms.

\section{Data availability}
The data that support the findings of this study are available from the corresponding author upon reasonable request.

\section{Acknowledgements}
We acknowledge fruitful discussions with John Close, Chris Freier, Kyle Hardman, Joseph Hope, and Paul Wigley, and insightful suggestions made by Franck Pereira dos Santos on behalf of the Atom Interferometry and Inertial Sensors team at SYRTE. S.S.S. was supported by an Australian Research Council Discovery Early Career Researcher Award (DECRA), Project No. DE200100495. O.H. was supported by IST Austria.

\bibliography{APL_perspective_bib.bib}

\end{document}